# PROSPECTS FOR A MUON SPIN RESONANCE FACILITY IN THE FERMILAB MUCOOL TEST AREA*

John A. Johnstone[†], Carol Johnstone[‡], Fermilab, Batavia, United States

## Abstract

This paper investigates the feasibility of re-purposing the MuCool Test Area (MTA) beamline and experimental hall to support a Muon Spin Resonance (MuSR) facility, which would make it the only such facility in the US. This report reviews the basic muon production concepts studied and operationally implemented at TRIUMF, PSI, and RAL and their application in the context of the MTA facility. Two scenarios were determined feasible. One, an initial minimal-shielding and capital-cost investment stage with a single secondary muon beamline that utilizes an existing high-intensity beam absorber and, another, upgraded stage, that implements an optimized production target pile, a proximate high-intensity absorber, and optimized secondary muon lines. A unique approach is proposed which chops or strips a macropulse of H- beam into a micropulse substructure – a muon creation timing scheme – which allows Muon Spin Resonance experiments in a linac environment. With this timing scheme, and attention to target design and secondary beam collection, the MTA can host enabling and competitive Muon Spin Resonance experiments.

## MUON SPIN RESONANCE

Muon Spin Resonance involves embedding polarized, positively charged muons in a sample material which allows the properties of the material to be measured and characterized at the microscopic atomic-structure level. The muons produced must be of sufficiently low energy that they stop within a few millimeters of material. Intense muon beams are currently produced first through bombardment of a low-Z production target with protons. The pions produced in nuclear reactions then decay "at rest" near the surface of the production target into low energy muons with polarization as high as 100% (a $\pi^+$ at rest decays into a 4.120 MeV, or 29.792 MeV/c, $\mu^+$ and a neutrino). The polarized muons are transported via a low-energy beamline to the sample material. The spin of the muon couples to the local magnetic field of the material, making them sensitive probes of the magnetic environment. The $\mu^+$ decays within a short time ($\tau_\mu = 2.197$ μs) to a positron and two neutrinos, with the positron emitted preferentially in the direction of the initial $\mu^+$ spin. Detection and measurements of the positrons reveals evolution of the $\mu^+$ polarization in the sample material, from which the properties of the target material can be deduced.

## A MUON SPIN RESONANCE FACILITY USING THE FERMILAB LINAC AND MUCOOL TEST AREA

Table 1: Linac full capability beam parameters. Note that the linac beam repetition rate is 200 MHz, but is accelerated with an 800 MHz RF system in every 4th RF bucket.

| Parameter | Value | Unit |
|---|---|---|
| Kinetic Energy | 401.5 | MeV |
| Energy Spread | 1 | MeV |
| RF Structure | 201.24 | MHz |
| Bunch Length | 0.05 | ns |
| Max Pulse Length | 80 | μs |
| Max Particles Per Bunch (28 mA) | $0.88 \times 10^9$ | |
| Max Particles Per Pulse | $1.4 \times 10^{13}$ | |
| Peak Current | 28 | mA |
| Avg Current | 34 | μA |
| Max Beam Power | 13.4 | kW |
| Beam Emittance (99%) | 8 | π mm-mrad |

### MTA Overview

The MTA experimental facility is supported by a primary beamline that extracts, transports, and delivers 400-MeV H- beam directly from the linac through a 12' shield wall and into a "stub beamline" which then opens into a 40' long, 20' wide hall. General Linac parameters are listed in Table 1, and Figure 1 shows the MTA exterior view and internal layout from the shield wall to the downstream high intensity dump.

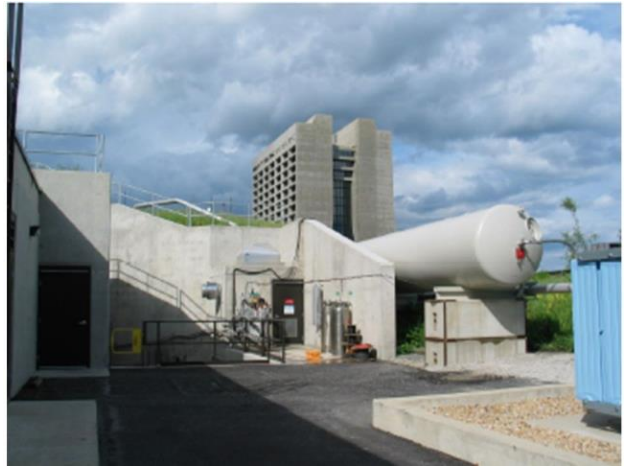

___________

* Work supported by Fermi Research Alliance, LLC under contract no. DE-AC02-07CH11359
† jjohnstone@fnal.gov
‡ cjj@fnal.gov

This manuscript has been authored by Fermi Research Alliance, LLC under Contract No. DE-AC02-07CH11359 with the U.S. Department of Energy, Office of Science, Office of High Energy Physics.

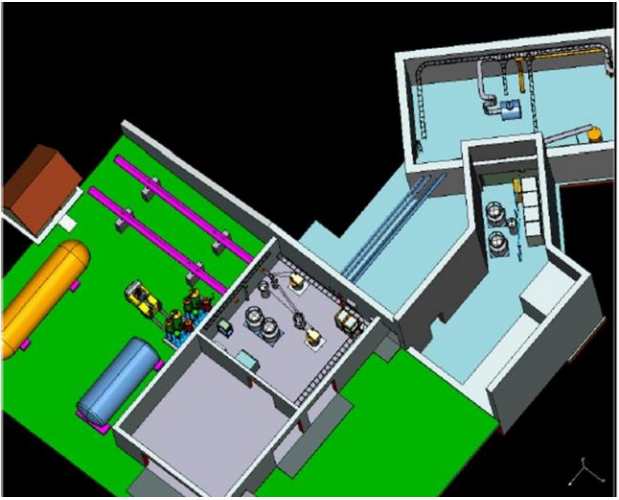

Figure 1: MTA facility and experimental hall.

*Scenarios for MuSR at MTA*

The ISIS nominal beam current is 200 µA at 800 MeV (1.25 x $10^{15}$ p/s). However, only a small fraction of this beam (≈4% or 5 x $10^{13}$ p/s) is available for µSR studies. A proton transmission rate of ~96% and low multiple scattering must be maintained to ensure that the neutron program is not significantly impacted. Such a constraint is not immediately applicable to a MuSR facility at MTA where with additional shielding the primary intensity could potentially reach 24 x $10^{13}$ p/s at full Linac (80µsec) capability; this would be equivalent to ISIS muon production for a 24% IL Be target, for example. However, this is without a muon creation or timing information needed for MuSR experiments but is very useful for target and secondary beamline optimization.

With additional local target shielding the high intensity absorber in the berm at the exit of the hall can be used for full intensity 15Hz Linac beam, or 60,000 pulses/hour.
- This represents an immediate, low cost approach to initiate a MuSR research and development facility.
- Full 15 Hz Linac intensity is available during synchrotron down or maintenance times.

A more ambitious option would convert the MTA experimental hall into a target hall with a target pile closely followed by a high intensity absorber.
- An enhanced muon production target [20% interaction length (IL) or higher] would permit increased multiple scattering highly increasing pion production

## SURFACE MUON PRODUCTION

Simulations of muon yields as a function of incident proton energy from threshold to 3 GeV have been studied for pions produced at 30 MeV/c or less for $10^9$ protons on target (POT) [1]. The results are displayed in Figure 2. There is a clear peak at the optimal energy of 500 MeV. The relative yield at MTA is also indicated. This is 80-85% of the TRIUMF value and, encouragingly, equal to the ISIS 800 MeV result.

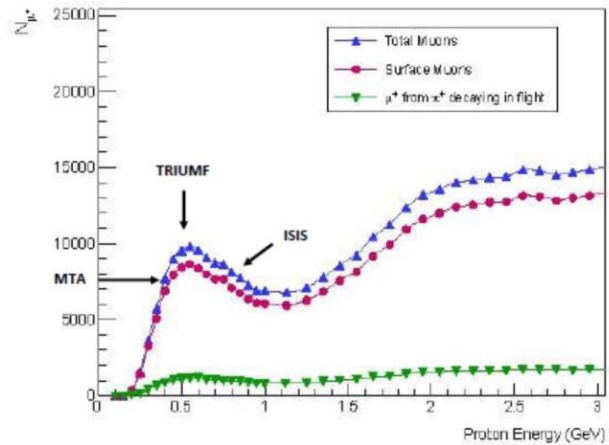

Figure 2: Variation of $\mu^+$ yield with incident proton energy for muons with momenta ≤30 MeV/c; i.e. surface muons [1].

## COMPARISON WITH ISIS AND TRIUMF

ISIS uses a 70 nsec pulse, 2.5x$10^{13}$ p/pulse, @40 Hz on a 4% IL target for MuSR (1 out of every 5 pulses is sent to a different target station),
- MuSR experiments at ISIS, however, run for only ~ half an hour which makes the integrated POT achievable using a high-intensity H- linac.

Therefore, for such short experiments, a more appropriate figure of merit is POT.
- The figure of merit for ISIS is an effective 7.2x$10^{16}$ POT ($10^{15}$ p/sec or 4x$10^{13}$ effective p/sec after multiplying by the 4% IL target).
- For Fermilab Linac operation at 28 mA (current capability of RFQ), 15 Hz, and a 25% IL target, this corresponds to a 61 microsec Linac pulse length for half an hour of dedicated beam – although without timing information for the surface muon.

*Secondary Beamline & Target Design*

Surface muon production is isotropic, as expected from a pion decaying at rest, which suggests that collection of backward muons is preferable for transporting a clean $\mu^+$ beam. Isotropic production is a powerful tool in that it allows the prospect of multiple secondary muon lines viewing the same target from different angles. At PSI this concept is implemented where 7 beamlines access two production targets. ISIS has 2 beamlines at 90º to the incident beam – one on opposite sides of the target. The acceptance of the beamline must also be considered, which is 5% for TRIUMF, but only 0.5% (0.071Ω) for ISIS, likely due to the line being multipurpose. This would further reduce time or pulse length estimates by an order of magnitude assuming the former hour of dedicated beam.

Secondary beamline acceptance and orientation must be folded into the evaluation of target details to make a preliminary recommendation of target thickness and configuration for a feasible MuSR implementation at MTA. Researchers from ISIS have demonstrated conclusively that target & beamline design are intimately inter-related issues [2]. In the ISIS study, to maximize the surface to volume ratio, they simulated splitting their graphite target into slices and studied the impact of the number and separation of slices and orientation with the beam. Within the existing ISIS extraction and transport configuration simple target modifications predicted double the detected surface muon yield. If the primary proton transmission is not constrained as it was in the ISIS study, heavier targets can be considered.

At MTA the target can be extended to a thicker sample of two or three slices. A target orientation angle of 45° was shown to increase muon yield without compromising collection from both sides of the target. Based on the extensive ISIS work, a two-slice configuration with this orientation yields a factor of 2 from two slices and a total factor of 2.8 from increased target thickness. The net gain in muon yield could potentially exceed that of an ISIS beamline – but again without timing information.

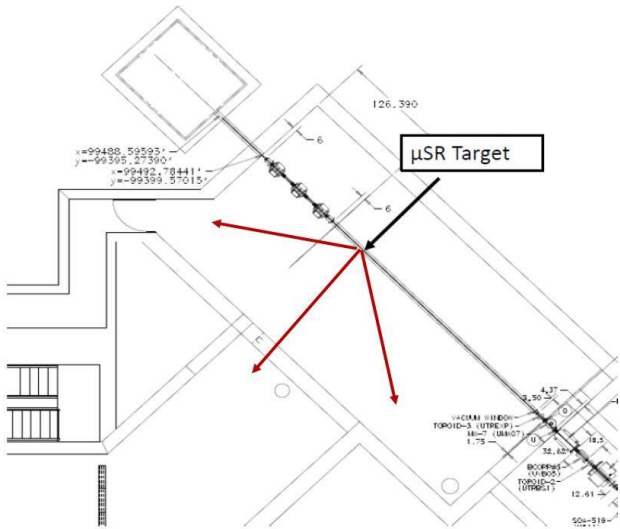

Figure 3: Proposed MuSR beamline components & possible transport directions (beam direction is bottom right to upper left).

*Experimental Surface Muon Timing Trigger*

For the experiment a unique creation timing trigger is required.
- ISIS 70 nsec pulse and 50 Hz cycle rate naturally allows a muon creation timing trigger
- TRIUMF is a CW beam with a 23 MHz RF structure (PSI is 51 MHz); bunches are too closely spaced for MuSR experimental timing. The primary beam and surface muon intensity is lowered until only one surface muon is incident on the ½ cm experimental target within a 10 μsec window.
- To decrease the probability of two muons produced during the 10 μsec window, a rate of one muon every 0.05 μsec is required. This corresponds to only 50,000 counts/sec of surface muons and the target TRIUMF uses is a solid (not sliced) 24% IL Be target.

To replicate ISIS operation using the Fermilab Linac, a beam pulse followed by two muon lifetimes of no beam is assumed; i.e. a 70 nsec batch of RF bunches separated by 4.4 μsecs of no beam formed by stripping the H$^-$ in the low 750- keV region where the current laser notcher operates.
- For an 80 μsec Linac pulse (max klystron pulse length), 18 70-nsec batches of RF bunches at 15 Hz containing $0.88 \times 10^9$ protons per RF bunch (28 mA) are formed.
- 4.4 μsec of no beam yields $2.4 \times 10^{11}$ p/linac pulse
- At 15 Hz, this corresponds to $3.6 \times 10^{12}$ p/sec, or $8.9 \times 10^{11}$ effective p/sec for a 25% IL target
- To achieve an effective $7.2 \times 10^{16}$ POT requires 22.5 hours of full Linac 15 Hz operation at an 80 msec pulse length for an exact duplication of ISIS conditions which are not optimized for MuSR beam.
- If the larger admittance of the TRIUMF beamline is used instead of the multipurpose ISIS muon line transmission, only 2.25 hours is required. A 2-slice target would reduce this time further to 1.125 hours.

## CONCLUSIONS

The decommissioned Fermilab MuCool Test Area has the potential to be re-purposed into a world-class Muon Spin Resonance Facility. A near-term, limited facilitation would require only modest upgrades to the MTA experimental hall and the installation of (existing) magnetic components. The hall itself can accommodate a high-intensity muon production target so a longer-term, more ambitious commitment to a dedicated MuSR facility into one competitive with existing facilities requires the experimental hall be converted to a target hall. Under these conditions, and an optimized target and secondary beamline would allow a MuSR experiment to be completed in a couple of hours.